\documentclass[
superscriptaddress,
 amsmath,amssymb,
 aps,
prfluids, 
]{revtex4-2}
\usepackage{xcolor}
\usepackage{graphicx}
\usepackage{dcolumn}
\usepackage{bm}
\usepackage{subeqnarray}
\usepackage[normalem]{ulem}
\usepackage{hyperref}
\hypersetup{
    colorlinks=true,
    linkcolor=blue,
    filecolor=blue,      
    urlcolor=blue,
    citecolor=blue
    }
\begin{document}

\title{Size Amplification of Jet Drops due to Insoluble Surfactants}

\author{Jun Eshima}
\thanks{Author contributions: J.E. and T. A. contributed equally to this work}
\affiliation{Department of Mechanical and Aerospace Engineering, Princeton University, Princeton, New Jersey 08544, USA}
\author{Tristan Aurégan}
\thanks{Author contributions: J.E. and T. A. contributed equally to this work}
\affiliation{Department of Mechanical and Aerospace Engineering, Princeton University, Princeton, New Jersey 08544, USA}
\author{Palas Kumar Farsoiya}
\affiliation{Department of Chemical Engineering, Indian Institute of Technology Roorkee,
Roorkee, Uttarakhand 247667, India}
\author{Stéphane Popinet}
\affiliation{Sorbonne Université, CNRS, UMR 7190, Institut Jean le Rond ’Alembert, F-75005 Paris, France}
\author{Howard A. Stone}
\affiliation{Department of Mechanical and Aerospace Engineering, Princeton University, Princeton, New Jersey 08544, USA}
\author{Luc Deike}
\email{ldeike@princeton.edu}
\affiliation{Department of Mechanical and Aerospace Engineering, Princeton University, Princeton, New Jersey 08544, USA}
\affiliation{High Meadows Environmental Institute, Princeton University, Princeton, New Jersey 08544, USA}

\begin{abstract}
Surface bubbles in the environment or engineering configurations, such as the ocean-atmosphere interface, sparkling wine, or during volcanic eruptions typically live on contaminated surfaces. A particularly common type of contamination is surface active agents (surfactants). We consider the effect of insoluble surfactant on jet drop formation by bubble bursting. Contrary to the observed trend that surfactants decrease the ejected drop radius for bubbles with precursor capillary waves, we find that surfactants increase the ejected drop radius for bubbles without precursor capillary waves - a regime characteristic of small bubbles. Consequently, the results have fundamental implications for understanding aerosol distributions in contaminated conditions. We find that the trend reversal is due to the effect of Marangoni stresses on the focusing of the collapsing cavity. We demonstrate quantitative agreement on the jet velocity and drop size between laboratory experiments and numerical simulations by using the measured surface tension dependence on surfactant concentration as the equation of state for the simulations. 
\end{abstract}

\maketitle

\section{Introduction}
When bubbles burst at a liquid-air interface, the resulting cavity collapses and the associated capillary waves focus to emit a liquid jet. The jet can then pinch to form drops, referred to as \textit{jet drops}, which contribute to ocean aerosol emissions \cite{veron_ocean_2015, deike_mass_2022}, health \cite{blanchard_ejection_1989, bourouiba_fluid_2021}, and industrial aroma \cite{ghabache_physics_2014}. Many applications involve complex solutions such as fluids with various rheological properties \citep{rudolph2024bubble,dixit_viscoelastic_2025}, contaminated interfaces with oil \cite{yang_jet_2025, yang_enhanced_2023a, sampath_aerosolization_2019}, or surface-active agents (surfactants) \citep{constante_dynamics_2021, pierre_influence_2022, pico_surfactantladen_2024, vega_influence_2024}.

\begin{figure*}
\includegraphics[width=\textwidth]{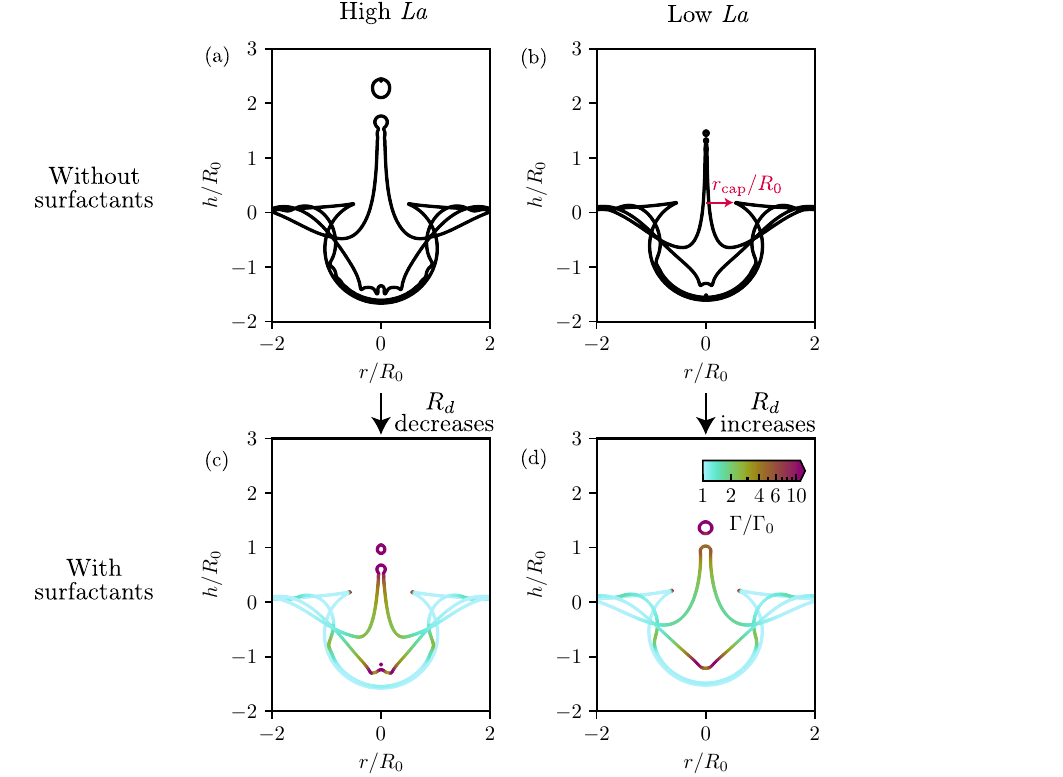} 
\caption{\label{fig:clean_contaminated} Jet drop formation without surfactants (a,b) and with surfactants (c,d), showing the effects of the Laplace number $\textit{La}=\rho \gamma_c R_0/\mu^2$. At high $\textit{La}$, the addition of surfactants increases the ejected drop radius $R_d$. At low $\textit{La}$, the addition of surfactants decreases $R_d$. Results shown are numerical. The axes are given by the interface profile $h$ and the radial coordinate is denoted $r$. The colors in (c,d) show the surfactant surface concentration $\Gamma$, nondimensionalized by the initial surfactant concentration $\Gamma_0$. (a) Spatial and temporal evolution shown for $\textit{La}=75000$, showing precursor capillary waves (ripples) as the jet is formed. Times shown are $t/t_{\text{ic}}=0,0.2,0.35, 0.7$ (colored light gray to black), where $t_{ic}=\sqrt{\rho R_0^3/\gamma_c}$ is the inertia-capillary timescale. (b) Spatial and temporal evolution shown for $\textit{La}=2400$, where the lack of precursor capillary waves allows the cavity to focus more efficiently. Times shown are $t/t_{\text{ic}}=0,0.2,0.39, 0.5$. (c) Spatial and temporal evolution shown for a typical case with surfactants for $\textit{La}=75000$. Times shown are $t/t_{\text{ic}}=0,0.2,0.43, 0.6$. For reference (to be defined later in the text), the surfactant parameters shown in (c) are given by $(\beta, \Delta \gamma_{\infty}, E)=(0.3, 0.52, 0.5)$. (d) Spatial and temporal evolution shown for a typical case with surfactants for $\textit{La} = 2400$. Times shown are $t/t_{\text{ic}}=0,0.2,0.45, 0.8$. The surfactant parameters shown in (c) are given by $(\beta, \Delta \gamma_{\infty}, E)=(0.3, 0.55, 0.5)$.
The Bond number shown in (a,c) is $\textit{Bo}= \rho g R_0^2/\gamma_c=0.13$ and (b,d) is $\textit{Bo}= \rho g R_0^2/\gamma_c=0.16$. The initial bubble shape is given by the Young-Laplace equations with the bubble cap removed at the foot of the cap $r= r_{\text{cap}}$.}
\end{figure*}

Jet drop production has been studied extensively  \cite{woodcock_giant_1953,blanchard_bursting_1954,duchemin_jet_2002}. 
The bursting of a bubble of radius $R_0$ in a Newtonian liquid of density $\rho$, dynamic viscosity $\mu$, with constant surface tension $\gamma_c$, and gravitational constant $g$ can be described by two non-dimensional numbers (assuming that the external air flow is negligible with high liquid-to-gas density and viscosity ratios),
\begin{equation}
    \textit{La}=\frac{\rho \gamma_c R_0}{\mu^2}, ~\textit{Bo}=\frac{\rho g R_0^2}{\gamma_c}\label{eq:LaBo},
\end{equation}
where the Laplace number $\textit{La}$ denotes a balance between capillary and viscous forces, and the Bond number $\textit{Bo}$ denotes a balance between  gravitational and capillary forces. The Laplace number is related to the Ohnesorge number $\textit{Oh}$ via $\textit{La}= \textit{Oh}^{-2}$ and can be considered a Reynolds number. It is known that for $\textit{La}\gtrsim \textit{La}_*\approx 500$, jet drops are ejected \cite{lee_size_2011,walls_jet_2015,brasz_minimum_2018,deike_dynamics_2018, berny_role_2020}. The jetting dynamics are controlled mainly by $\textit{La}$, with $\textit{Bo}$ controlling the initial bubble shape and modifying the jetting dynamics for $\textit{Bo}  \gtrsim 0.1$ \cite{ganan-calvo_revision_2017, deike_dynamics_2018,ganan-calvo_scaling_2018, gordillo_capillary_2019,blanco_sea_2020, berny_role_2020}. For $\textit{La} > \textit{O}(10^4)$, the capillary waves due to the collapsing cavity contain many precursor waves (ripples) (Fig.~\ref{fig:clean_contaminated}(a)), whereas for $\textit{La}=\textit{O}(10^3)$, the precursor waves are damped by viscosity and the collapsing cavity can focus more efficiently (Fig.~\ref{fig:clean_contaminated}(b)), producing self-similar dynamics that give a faster and narrower jet \cite{zeff_singularity_00,duchemin_jet_2002, ghabache_physics_2014, lai_bubble_2018}, leading to smaller jet drops. For water surface bubbles, bubbles with $\textit{La}=\textit{O}(10^3)$ correspond to bubble radii $R_0 \approx 10$ - $ 100$ \textmu  m, which is a size range commonly observed in surface bubble distributions arising in nature and engineering applications, such as on the ocean surface \cite{deike_mass_2022, deike_mechanistic_2022, mazzatenta_linking_2025, wang_role_2017, blanco_sea_2020, ganan_ocean_2023,jiang_submicron_2022}. Thus, understanding the jet drop production of such bubbles is important in understanding aerosol distributions from bubble bursting.

In practice, interfaces are often contaminated with surfactants from biological or anthropogenic origin. The surfactant surface concentration $\Gamma$ affects the surface tension of the interface via some equation of state $\gamma = \gamma(\Gamma)$ (often referred to as the surface tension isotherm) \cite{manikantan_surfactant_2020}. The resulting surface tension gradient gives rise to a stress: the Marangoni stress. 
Jet drop production depends significantly on the surface tension and therefore is affected by surfactants, as shown experimentally and numerically \cite{constante_dynamics_2021,pierre_influence_2022, neel_role_2022, vega_influence_2024, pico_surfactantladen_2024}. Previous studies \cite{constante_dynamics_2021,pierre_influence_2022, vega_influence_2024, pico_surfactantladen_2024} investigated surfactant effects for $\textit{La}\in [2\cdot10^4, 10^5]$, where precursor capillary waves are present, and found that adding surfactants decreases the ejected drop radii. Indeed, at high $\textit{La}$, surfactants allow the cavity to focus more since the Marangoni stress damps the precursor waves, leading to a faster and narrower jet (see Fig.~\ref{fig:clean_contaminated}(c) for a typical case with surfactants at high $\textit{La}$). 

Quantitative comparisons between experiments and numerical simulations have remained limited \cite{liao_deformation_2006} as experiments report the surfactant bulk concentration without having access to spatial and temporal variations in surface concentration and the surfactant equation of state is not always available. Separately, two-phase Navier-Stokes simulations with surfactant transport and Marangoni stresses 
enable detailed access to the surfactant dynamics.
However, connecting the multiple numerical parameters influencing Marangoni stresses to practical surfactant effects has remained challenging. Experiments typically report a bulk concentration, which is connected to the Marangoni stresses through an equation of state that is not always known. Moreover, real surfactants typically have many chemical and physical properties beyond what is typically considered in theoretical models or a tabulated equation of state for ideal/pure surfactants. Further, even if the theoretical properties are well-understood, making quantitative connections to experimental measurements remains difficult. For example, it is difficult to know how much of the added surfactant ultimately resides on the surface, or what are the adsorption kinetics. As result, the previous studies on jet drops with surfactants were either strictly experimental \cite{pierre_influence_2022, vega_influence_2024} or strictly numerical \cite{constante_dynamics_2021,pico_surfactantladen_2024}. 

In this paper, we show, through experiments and simulations, that for bubbles with efficient self-similar cavity collapse that is not controlled by precursor capillary waves ($\textit{La}=\textit{O}(10^3)$), the ejected drop radii increase upon the addition of insoluble surfactants by changing the shape of the self-similar collapsing cavity (see Fig.~\ref{fig:clean_contaminated}(d) for a typical case with surfactants at low $\textit{La}$). 
We present a systematic approach to quantitatively relate experiments and simulations. We use experimentally measured surface tension isotherms $\gamma = \gamma(\Gamma)$ as the equation of state in the simulation, and obtain excellent agreement between the experimental and simulation results in terms of the temporal and spatial evolution of the cavity collapse, jet formation, and drop ejection.

The structure of the paper is as follows: we first describe, and then quantitatively compare the numerical and experimental frameworks. Then, the effects of surfactants on jet drop production at low $\textit{La}$ are described. Finally, we propose a physical mechanism for the trend reversal, dependent on $\textit{La}$ and based on how Marangoni stresses modulate the cavity collapse.

\section{Methods}

Our numerical simulations solve the 2D axisymmetric two-phase Navier-Stokes equations using the coupled level-set/volume-of-fluid method \cite{sussman_coupled_2000,abu_conservative_2018} 
using the Basilisk open-source library \cite{popinet_accurate_2009,Basilisk}. Air-water density and viscosity ratios are used. The numerical configuration is adapted from Berny et al. \cite{berny_role_2020,berny_statistics_2021}, who considered bubble bursting without surfactants. The Marangoni stress is resolved with an integral formulation for the surface tension \cite{abu_conservative_2018, saini_implementation_2025} and the insoluble surfactant concentration is transported via a coupled phase-field/volume-of-fluid method \cite{farsoiya_coupled_2024, jain_model_2024} (see supplementary material). The computational domain is given by $(r,z) \in [0,10R_0] \times [-5R_0,5R_0]$, with $z=0$ defining the flat surface away from the bubble. We use adaptive mesh refinement with level 12 (i.e., the smallest grid size is the size of the box $10R_0$ divided by $2^{12}$, leading to $\approx 800$ points across the bubble diameter), with grid convergence confirmed by comparison with level 13 simulations (see supplementary material).

Just after bubble bursting, the liquid-air interface is initialized using the Young-Laplace equations \cite{toba_drop_1959, princen_shape_1963, bubbleShape_h} to compute the theoretical shape of the cavity, with the bubble cap removed at the foot of the cap $r = r_{\text{cap}}$ (see Fig.~\ref{fig:clean_contaminated}) \cite{duchemin_jet_2002, ghabache_physics_2014,berny_role_2020, constante_dynamics_2021, pico_surfactantladen_2024}. We consider a uniform initial surfactant concentration $\Gamma = \Gamma_0$, and test the sensitivity to surfactant accumulation on the cap film, which retracts into the tip at $r = r_{\text{cap}}$ upon bursting, represented by an initial non-uniform concentration
\begin{equation}
    \Gamma(s_{\text{cap}}) = \Gamma_0 \left(1+ER_0 \delta(s_{\text{cap}})\right), \label{eq:initial_cond}
\end{equation}
where $s_{\text{cap}}$ is the arclength from the tip at $r = r_{\text{cap}}$ and $\delta(s_{\text{cap}})=\sigma^{-1}(2\pi)^{-\frac{1}{2}}e^{-s^2/(2\sigma^2)}$ is the unit Gaussian distribution with standard deviation $\sigma$. The non-dimensional parameter $E$ sets the total amount of excess surfactants, which is approximately $2 \pi r_{\text{cap}}\int_{-\infty}^{\infty}\Gamma_0 ER_0 \delta(s_{\text{cap}}) ds_{\text{cap}}=2 \pi r_{\text{cap}} \Gamma_0 ER_0$, independent of $\sigma$. Then, the dynamics are also independent of $\sigma$ for $\sigma/R_0$ sufficiently small (see supplementary material). A model to evaluate $E$ is to assume that the surfactant that was in the cap film, approximately $2\pi r_{\text{cap}}^2\Gamma_0$ (since the total surface area of the two-sided cap is $2 \pi r_{\text{cap}}^2$ for small $\textit{Bo}$), ends up solely in the tip. Equating with the total surfactant excess $2 \pi r_{\text{cap}}\Gamma_0 ER_0$ gives $E \approx r_{\text{cap}}/R_0$. 
Although our estimate is simple, it models the complexity of the film retraction process \cite{constante_role_2022,shaw_film_2024} and possible surfactant enrichment in the cap film due to complex drainage processes \cite{poulain_ageing_2018, poulain_biosurfactants_2018}.

Our experiments are performed by generating bubbles in a small tank and recording their bursting at the surface; see previous experiments with similar setups \citep{ghabache_physics_2014,pierre_influence_2022,ji_secondary_2023a}. The bubble lifetimes vary from $0-16$ s. The bubbles are generated using a syringe pump pushing air through a small capillary, resulting in sizes $R_0 \approx 1.0$~mm. We use three different solutions (see table \ref{tab:solvent}), one consisting only of deionized water, one with 33\% by weight of glycerol and one with 50\% glycerol. For each of the solutions, various amounts of Triton X-100, a nonionic surfactant, are added, in concentrations ranging from 1 to 20\% of the critical micelle concentration (CMC). The CMC is defined as the bulk concentration above which micelles start forming and the surface tension no longer depends on concentration. For Triton X-100 this concentration is 230 \textmu mol/L \citep{fainerman_adsorption_2009}. The surfactant adsorbs or desorbs from an interface slowly: the typical timescale ranges from minutes to hours \citep{fainerman_adsorption_2009,fainerman_adsorption_2009a,miller_dynamic_2017,erinin_effects_2023}, allowing us to assume insolubility on capillary timescales ($O(1~{\rm ms})$).

\begin{table}
\setlength{\tabcolsep}{9pt}
\begin{tabular}{c|cccc}
 wt \% &$\textit{La}$ &$\textit{Bo}$ & $\Delta \gamma_{\infty}$ & $\beta$\\
\hline
  50 & 2\,000 - 2\,700 & 0.15 - 0.17& 0.55 &0.1 - 0.3\\
  33 & 8\,800 - 12\,000& 0.12 - 0.16 & 0.58 & 0.1 - 0.4\\
  0 & 67\,000 - 83\,000& 0.12 - 0.14& 0.52 & 0.0 - 0.4\\
\end{tabular}
\caption{\label{tab:solvent}%
The non-dimensional parameters in the experiment, for the three glycerol weight fractions used \cite{cheng_formula_2008, takamura_physical_2012}. The uncertainty of $\textit{La}$ is mostly due to the temperature (estimated as $21\pm1^\circ$C), which affects the dynamic viscosity $\mu$. See supplementary material for the dimensional parameters.} 
\end{table}

To quantitatively compare experiments and simulations in addition to the trends, we measure the surface tension isotherm (equation of state) experimentally, which is used into the simulations. In a Langmuir trough, the surface tension is measured using the Wilhelmy plate method, where two barriers compress the surfactant-covered surface, reducing the area $A$ available for surfactant molecules. The surfactant is considered insoluble during the measurement (again, the adsorption and desorption processes are slow \citep{erinin_effects_2023}), and a change in surface area therefore corresponds to a change in surface concentration $\Gamma$ through $A / A_0 = \Gamma_0 / \Gamma$, where $A_0$ and $\Gamma_0$ are the area and surface concentration in the initial state, respectively. Curves obtained through this procedure are shown (markers) in Fig. \ref{fig:intro}(a).  

We fit the isotherms with a model that takes into account the surfactant physics while minimizing the number of parameters (dashed curves in Fig. \ref{fig:intro}(a)). 
In particular, we model the surface tension saturation that is experimentally visible at large $\Gamma$. Such saturation is believed to be due to rearrangements of surfactants on the interface \citep{kaganer_structure_1999}. The  saturation is important for the jet drop dynamics since the focusing of the capillary waves during cavity collapse leads to a much larger surfactant concentration at jet formation than the initial concentration (see Fig.~\ref{fig:clean_contaminated}(c,d)). For simplicity, we use a two-parameter $\tanh$ model \cite{liao_deformation_2006, farsoiya_coupled_2024}:
\begin{equation}
    \frac{\gamma}{\gamma_c} = 1 - \Delta \gamma_{\infty}\tanh\left(\frac{\beta}{\Delta \gamma_{\infty}} \frac{\Gamma}{\Gamma_0}\right) \label{eq:non_dim_isotherm}
\end{equation}
with two nondimensional parameters $\Delta \gamma_\infty$ and $\beta$.

Physically, $\Delta\gamma_\infty$ is the nondimensional difference in surface tension between a clean and saturated interface since (\ref{eq:non_dim_isotherm}) gives $\gamma / \gamma_c=1$ at $\Gamma = 0$ and $\gamma / \gamma_c\approx 1-\Delta \gamma_{\infty}$ for $\Gamma \gg \Gamma_0$. Therefore, $\Delta\gamma_\infty$ is constant for a given solution. In this paper, only the experimentally-derived $\Delta \gamma_{\infty}$ is used, but the simulations show that there is sensitivity to $\Delta \gamma_{\infty}$ (see supplementary material).
The surfactant parameter $\beta$ can be interpreted as the slope of the isotherm since $\gamma / \gamma_c\approx 1-\beta (\Gamma/\Gamma_0)$ for $\Gamma \ll \Gamma_0$. Alternatively, $\beta$ is a nondimensional surfactant parameter that combines the initial surface surfactant concentration and the strength of the surfactant (the dimensional version of Eq.~3 has the term $\tanh(\tilde{\beta}\Gamma)$ with $\beta = \Delta \gamma_{\infty}\tilde{\beta}\Gamma_0$). The fitted parameters are summarized in Table \ref{tab:solvent}; note that in Fig.~\ref{fig:intro}(a) the case with no surfactants has a non-zero slope, resulting in a non-zero $\beta$. This effect is likely due to the presence of trace amounts of surface active compounds in the case with deionized water and glycerol only.

The simulation is sensitive to the particular choice of the function used to model the two-parameter isotherm (\ref{eq:non_dim_isotherm}). While this ambiguity can be avoided by fitting the isotherm more closely using more parameters, which gives a comparable performance in describing the dataset, the use of more parameters leads to interpretability and overfitting issues (discussed in more details in the supplementary material).

\section{Results}

\begin{figure*}
\includegraphics[width=0.95\textwidth]{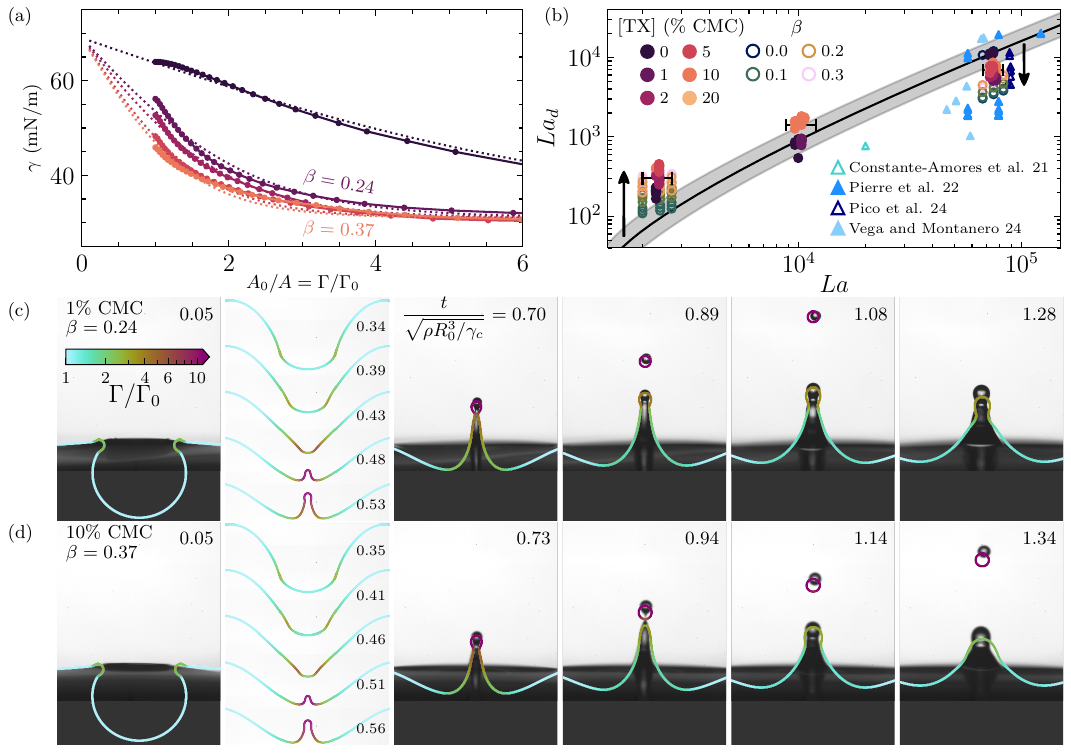} 
\caption{\label{fig:intro} Surfactant effects on jet drops. (a) Experimentally measured surfactant equation of state (markers) for several bulk surfactant concentrations as a function of the surface surfactant concentration $\Gamma$ (see legend in (b)), with the corresponding fits (dashed curves) according to Eq. \eqref{eq:non_dim_isotherm}. Glycerol weight fraction is 50\%.
(b) Laplace number $La_{d}=\rho \gamma_c R_d/\mu^2$ of the first ejected drop radius $R_d$ as a function of $\textit{La}$. Experimental points (filled) are 
colored by the bulk Triton X-100 concentration. Numerical points (open) 
are colored by the surfactant parameter $\beta$ (Eq. \eqref{eq:non_dim_isotherm}). 
The black line and shaded area represent the average and typical spread found in the literature data (without surfactant) \citep{berny_role_2020,brasz_minimum_2018}. Triangles represent experimental \citep{pierre_influence_2022, vega_influence_2024} (filled) and numerical \citep{constante_dynamics_2021, pico_surfactantladen_2024} (open) data available in the literature on jet drops with surfactant. Arrows indicate the effect of adding surfactant at a given $\textit{La}$.
(c,d) Illustrations of the jet drop dynamics and comparisons between experiments and simulations ((c,d): $(\textit{La},\textit{Bo}, \Delta \gamma_{\infty}, E) = (2000, 0.16, 0.55, 0.5)$, (c): $\beta=0.24$, (d): $\beta=0.37$). The background images are experimental snapshots and the color \cite{Crameri2018} overlays show the interface profile and the surfactant concentration from the simulations. Supplementary videos are available.
}
\end{figure*}

Using our dual approach combining experiments and numerics, we show the effect of insoluble surfactants on the size of the first ejected drop in Fig.~\ref{fig:intro}(b). We plot the Laplace number $\textit{La}_d = \rho \gamma_c R_d/\mu^2$ of the first ejected drop radius $R_d$ as a function of $\textit{La}$, where the experimental points are shown as filled symbols and the numerical points are shown as empty symbols. The black line and shaded area represent the average and typical spread found in the literature for the bubbles without surfactants. At low $\textit{La}$ ($\approx 2\,400$), we see up to a roughly four-fold increase of the drop radius due to the presence of surfactants, which is in stark contrast to the trend at high $\textit{La}$ ($\approx 70\,000$), where our data and all the previous works \citep{constante_dynamics_2021, pierre_influence_2022, pico_surfactantladen_2024, vega_influence_2024} show up to a roughly ten-fold decrease in the ejected drop radius due to the presence of surfactants. In between these two regimes ($La \approx 10\,000$) the experimental data shows little to no trend with surfactant concentration and is contained within the typical spread from previous works in clean conditions.
All of our data is therefore consistent with a robust reversal of the trend with surfactant concentration between low and high Laplace number regimes, which we also illustrated in figure 1.

The use of the experimentally measured isotherm as the equation of state in the numerics yields excellent agreement when comparing the temporal evolution of the cavity collapse, jet formation and drop ejection in the simulations and experiments. In Fig.~\ref{fig:intro}(c) and (d) we show the time evolution leading to jet drop formation, combining experiments and numerics for low Laplace number cases. These sequences are made with no fitting parameters: all physically relevant quantities are measured or from the literature, allowing us to link the observed time and space scalings between experiments and numerics. The origins of time are chosen following the time at which the jet crosses the still mean liquid level. The simulation data shows the dimensionless surfactant concentration on the interface and has access to information in the cavity. We can therefore see the collapse of the cavity in the second panel  (from the left) of (c,d), which shows the high concentration of surfactant in the high curvature areas near the bottom and then later in the jet. When the jet starts to be visible above the surface (see the panels for time $t/\sqrt{\rho R_0^3/\gamma_c}=0.70$ onwards in (c) and $t/\sqrt{\rho R_0^3/\gamma_c}=0.73$ onwards in (d)), we see excellent agreement in its length and width between experiment and simulation, and in particular the instant at which the first drop detaches is well captured. Finally, the simulation allows us to notice the high surfactant enrichment in the ejected drop: more than 10 times the initial surface concentration. 

Between Fig.~\ref{fig:intro}(c) and (d) only the initial surfactant surface concentration is changed. In the experiment, this is done by changing the bulk concentration of surfactants from 1\% of the CMC in (c) to 10\% in (d). Numerically, only $\beta$ is changed from 0.24 to 0.37. As a result, notice how the timescale until drop ejection has changed, and this change is captured by the simulation. In addition, the width of the jet, and therefore the size of the drop, has increased by a similar amount in experiments and numerics.

\begin{figure}
\includegraphics[width=\textwidth]{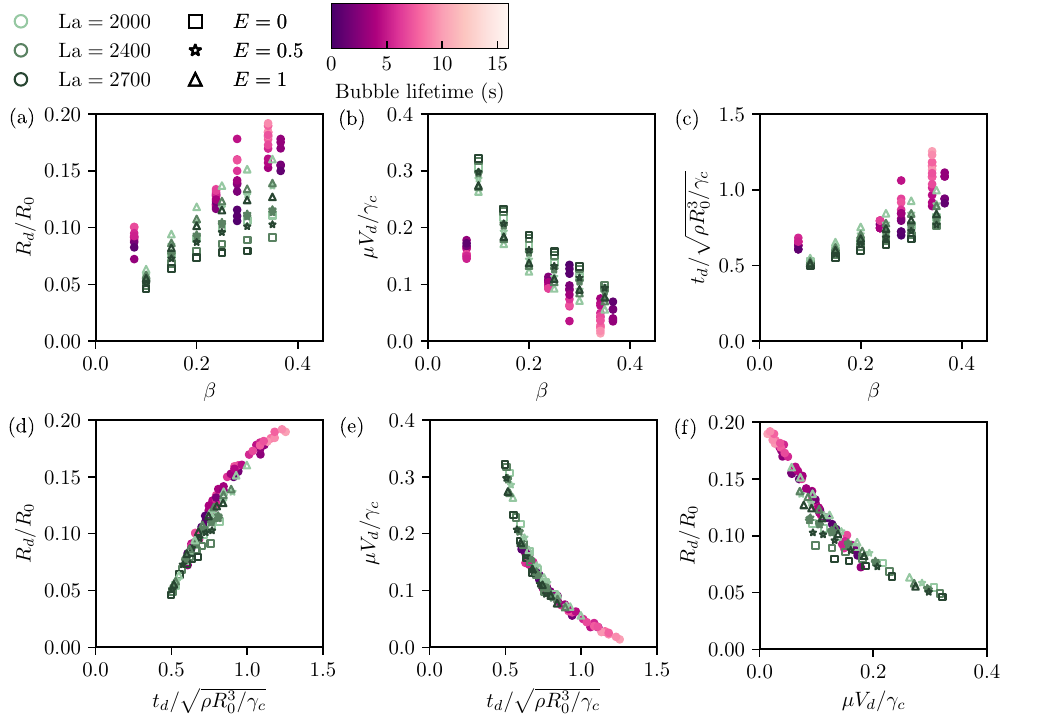} 
\caption{\label{fig:parameter_sweep} Effect of the surfactant parameter $\beta$ on the first ejected drop radius $R_d$, velocity $V_d$, and pinch-off time $t_d$. (a,b,c) Relationships between $(R_d, V_d, t_d)$ and $\beta$, nondimensionalized by the bubble radius $R_0$, capillary velocity $\gamma_c/\mu$, and inertiocapillary timescale $\sqrt{\rho R_0^3/\gamma_c}$ respectively. (d,e,f) Relationships between non-dimensionalized $(R_d, V_d, t_d)$. The same numerical and experimental data points are used throughout (a-f). For $\textit{Bo}=0.16$ as considered, $r_{\text{cap}}/R_0\approx 0.5$ and hence results with $E = 0-1$ are shown.
}
\end{figure}

We further investigate the effect of surfactants on the first ejected jet drop for low $\textit{La}$, systematically varying $\beta$ and comparing with the experimental case with 50\% glycerol. 
Sensitivity to $\textit{La}$ and non-uniformity in the initial surfactant distribution is considered so as to compare with the experimental variations in viscosity and bubble lifetime.
We focus on three dependent variables $(R_d, V_d, t_d)$ accessible both experimentally and numerically, where the first ejected drop radius is $R_d$, $V_d$ is the velocity of the first ejected drop (at pinch off), and $t_d$ is the time between bubble rupture and jet pinch off as a function of the surfactant parameter $\beta$.
In Fig.~\ref{fig:parameter_sweep} the experimental points (solid) are colored by the bubble lifetime prior to bubble rupture. The shape of the numerical points (open) are given by $E$ and are colored by $\textit{La}$. Fig.~\ref{fig:parameter_sweep}(a,b,c) show that as $\beta$ increases, the drop radius $R_d$ increases, drop velocity $V_d$ decreases, and pinch-off time $t_d$ increases for both experiments and simulations. 
Increasing $E$ or decreasing $La$ in the simulations results in $R_d$ and $t_d$ increasing and $V_d$ decreasing.
The experiment shows a spread for each measured $\beta$. Such a spread is physical and arises from the variation in bubble lifetime, whose effect seems to be similar to increasing the excess surfactant $E$, which suggests that variations in accumulation of surfactant at the top of the bubble, as captured by the parameter $E$, may be an explanation for lifetime effects \cite{poulain_ageing_2018, shaw_film_2024, auregan2025surface}.  
Indeed, when considering relationships implicit in $\beta$ (see Fig.~\ref{fig:parameter_sweep}(d,e,f)), the experimental and numerical points collapse onto a curve with strong agreement between the experimental and numerical data. Such agreement suggests that the jet drop production dynamics are being resolved accurately  by the simulation. Again, it should be noted that the exact degree of agreement between the simulation and experiment is sensitive to the particular model of the isotherm used. The numerical collapse onto a single curve also suggests the following. Changing either $\beta$ or $E$ moves the points along the same curve in Fig.~\ref{fig:parameter_sweep}(d,e,f), even though the surfactant distribution on the jet changes. If the primary effect of the surfactants on jet drop production was through the effects of Marangoni stresses on the pinching of the jet after the jet production, then one would not expect such a collapse, due to the differing surfactant concentration. This suggests that the primary effect of the surfactants on jet drop production at low $\textit{La}$ is through the modification of the cavity shape (more precisely, the capillary pressure).

We show that the size amplification due to the surfactants occurs since Marangoni stresses disturb the self-similar collapse of the cavity shape, leading to a slower and thicker jet formation. Indeed, without surfactants, at low $\textit{La}=1000-5000$ \cite{lai_bubble_2018}, it has been shown that the interface profile is self-similar just before and after jet production \cite{zeff_singularity_00, duchemin_jet_2002, ghabache_physics_2014,  lai_bubble_2018} at time $t_0$ with a $|t-t_0|^{\frac{2}{3}}$ scaling.

\begin{figure}
\includegraphics[width=\textwidth]{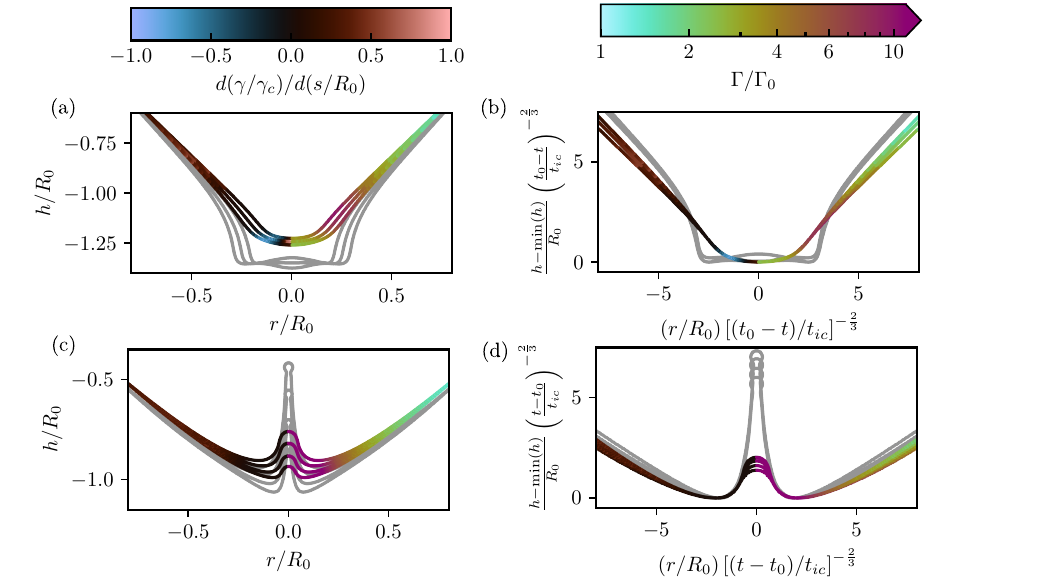} 
\caption{\label{fig:simil_soln} Self-similar profile comparison between a clean ($\textit{La} = 2400, \textit{Bo} = 0.16$) and correspondingly contaminated ($\textit{La} = 2400, \textit{Bo} = 0.16, \beta = 0.3, \Delta \gamma_{\infty}=0.55, E = 0.5$) bubble. The clean case is analogous to Figs.~3,4 of \cite{lai_bubble_2018}. The interface is given by $z=h(r,t)$. (a) The time evolution of the cavity collapse in time $(t < t_0)$ for the clean (gray) and contaminated bubble (colors). The times shown are $(t_0-t)/t_{ic}=0.018, 0.024, 0.03$, where $t_{ic}=\sqrt{\rho R_0^3/\gamma_c}$ is the inertia-capillary timescale. (b) The time evolution of the cavity collapse in self-similar coordinates, corresponding to curves in (a). (c) The time evolution of jet production ($t>t_0$) (same convention as (a)), at times $(t-t_0)/t_{ic}=0.008, 0.011, 0.015, 0.019$. (d) The time evolution of jet production in self-similar coordinates corresponding to curves in (c). In (b,d), $\min$ is taken over the profile of the cavity or jet. The colors for (a,b,c,d) are given by the surfactant concentration $\Gamma$ and Marangoni stress $d\gamma/ds$ for surface tension $\gamma$ and arclength $s$ measured from $r=0$. The particular choice of times plotted in (a,b,c,d) is the same as \cite{lai_bubble_2018}, which captures the self-similarity. 
}
\end{figure}

Fig.~\ref{fig:simil_soln} shows the cavity collapse ($t<t_0$) in physical (a) and self-similar (b) coordinates, and the jet production ($t>t_0$) in physical (c) and self-similar (d) coordinates. A typical clean bubble (gray) is compared to a corresponding contaminated bubble (colored) at low $\textit{La} = 2400$. The clean case in Fig.~\ref{fig:simil_soln} is the analogue to Figs.~3,4 of Lai et al. \cite{lai_bubble_2018} (\cite{lai_bubble_2018} did not consider surfactants). 

As noted in the literature, for clean bubbles, at low $\textit{La}$, the cavity forms a sharp corner prior to jet formation (see Fig.~\ref{fig:simil_soln}(a,b)), which results in high capillary pressure that leads to a fast and thin jet (see Fig.~\ref{fig:simil_soln}(c,d)). In comparison, as can be seen in Fig.~\ref{fig:simil_soln}(a,b), the case with surfactants has a smoother corner, which in combination with a lower surface tension due to the presence of surfactants, results in a smaller magnitude capillary pressure and leads to a slower and thicker jet (see Fig.~\ref{fig:simil_soln}(c,d)). Physically, there is an accumulation of surfactants (see Fig.~\ref{fig:simil_soln}(a,b)) near the corner of the cavity (such accumulation has also been seen in \cite{constante_dynamics_2021, pico_surfactantladen_2024} for high $\textit{La}$), which results in a low surface tension region leading to Marangoni stress that smooths the corner. Mathematically, it has been shown that the possible self-similar interface profile \cite{sierou_selfsimilar_2004} for inertio-capillary cavity collapse is a family parametrized by the far-field (with respect to the local region of cavity collapse) cone angle and interfacial flow field. Thus, the cavity shape can be interpreted as the result of Marangoni flow that acts to select a smoother corner. It could be valuable to consider the spatial and temporal evolution of the curvature in greater detail in the presence of surfactants (e.g., such as in \cite{gordillo_capillary_2019}).

Separately, for $\textit{La} \gg 5000$, multiple capillary waves appear as precursor waves ahead of the dominant wave (referred to as the corner above). In such a regime, the details of the precursor waves play a key role in the jet behavior, and hence the drop amplification mechanism identified in this paper, is not applicable. Indeed, previous studies have suggested that for $\textit{La} \gg 5000$, the Marangoni stress damps the capillary waves in such a way that the drop radius decreases \cite{constante_dynamics_2021, pierre_influence_2022, pico_surfactantladen_2024, vega_influence_2024}; which we also observe in our numerical and experimental data (see Fig.~\ref{fig:intro}(a) and supplementary material). 

\section{Conclusion}

In summary, we demonstrate that the trend reversal of drop radius with $\textit{La}$ (when adding surfactant), first an increase and then decrease, is due to the change in the dominant physical mechanism that produces the jet drop at low $\textit{La}$ (high capillary pressure of the cavity corner) and high $\textit{La}$ (precursor wave). We have identified that the Marangoni stress acts to increase the ejected drop radius for low $\textit{La}$, distinct from the high $\textit{La}$ regime, where the ejected drop radius decreases instead. These arguments are supported by the experimental observations (see Fig.~\ref{fig:intro}(a)), which show that the transition occurs near $\textit{La} \approx 10^4$. 

As a direct extension, it would be interesting to include how other physical effects, such as surfactant solubility and viscoelasticity, would affect the trend reversal. 
Considering a larger class of non-uniform initial surfactant distributions, may also be important. Finally, there are many problems where the details of Marangoni flow are important, including surfactant spreading \cite{gaver_dynamics_1990}, coating flows \cite{quere_fluid_1999}, bubble and droplet breakup, surface and breaking waves \cite{erinin_comparison_2023} Many Marangoni flow problems would likely benefit from the approach taken in this paper, where experimentally-derived surface tension isotherms are used in the numerical simulations.   

Rodríguez-Aparicio et al. \cite{rodriguez_critical_2025} have posted an independent study that reports experimental results at low $\textit{La}$ compatible with the discussion presented here.

The code for the main text is available online \cite{eshima_code_2025}. The numerical and experimental data for the main text is available online \cite{eshima_data_2025}.

\begin{acknowledgments}
\textcolor{black}{Contribution: Conceptualization: JE, TA, LD. Investigation: JE, TA, PKF, SP, HAS, LD; experiments were performed by TA, while numerical simulations were performed by JE. Formal Analysis: JE, TA, LD, PKF, HAS, SP. Software: JE, PKF, SP; the surfactant transport module was developed by PKF with the basilisk library developed by SP. Writing – Original Draft Preparation: JE, TA, LD. Writing – Review \& Editing: JE, TA, LD, PKF, SP, HAS.}
\\
This work was supported by NSF grant 2242512 to L.D. This work used TAMU Faster at Texas A\&M High Performance Research Computing through allocation OCE140023 to L.D. from the Advanced Cyberinfrastructure Coordination Ecosystem: Services \& Support (ACCESS) program \cite{boerner_access_2023}, which is supported by U.S. National Science Foundation grants 2138259, 2138286, 2138307, 2137603, and 2138296. 
\end{acknowledgments} 

\bibliography{jet_drops}

@article{abu_conservative_2018,
  title={A conservative and well-balanced surface tension model},
  author={Abu-Al-Saud, Moataz O and Popinet, St{\'e}phane and Tchelepi, Hamdi A},
  journal={Journal of Computational Physics},
  volume={371},
  pages={896--913},
  year={2018},
  publisher={Elsevier}
}

@article{sampath_aerosolization_2019,
  title={Aerosolization of crude oil-dispersant slicks due to bubble bursting},
  author={Sampath, Kaushik and Afshar-Mohajer, Nima and Chandrala, Lakshmana D and Heo, Won-Seok and Gilbert, Joshua and Austin, David and Koehler, Kirsten and Katz, Joseph},
  journal={Journal of Geophysical Research: Atmospheres},
  volume={124},
  number={10},
  pages={5555--5578},
  year={2019},
  publisher={Wiley Online Library}
}

@article{jain_model_2024,
  title={A model for transport of interface-confined scalars and insoluble surfactants in two-phase flows},
  author={Jain, Suhas S},
  journal={Journal of Computational Physics},
  volume={515},
  pages={113277},
  year={2024},
  publisher={Elsevier}
}

@incollection{boerner_access_2023,
  title={Access: Advancing innovation: Nsf’s advanced cyberinfrastructure coordination ecosystem: Services \& support},
  author={Boerner, Timothy J and Deems, Stephen and Furlani, Thomas R and Knuth, Shelley L and Towns, John},
  booktitle={Practice and Experience in Advanced Research Computing 2023: Computing for the Common Good},
  pages={173--176},
  year={2023}
}

@article{constante_role_2022,
  title={Role of surfactant-induced Marangoni stresses in retracting liquid sheets},
  author={Constante-Amores, Cristian Ricardo and Chergui, Jalel and Shin, Seungwon and Juric, Damir and Castrej{\'o}n-Pita, JR and Castrej{\'o}n-Pita, Alfonso Arturo},
  journal={Journal of Fluid Mechanics},
  volume={949},
  pages={A32},
  year={2022},
  publisher={Cambridge University Press}
}

@misc{Crameri2018,
  author       = {Fabio Crameri},
  title        = {Scientific colour maps},
  year         = {2018},
doi = {10.5281/zenodo.1243862},
  url          = {https://doi.org/10.5281/zenodo.1243862},
}

@article{cheng_formula_2008,
  title={Formula for the viscosity of a glycerol- water mixture},
  author={Cheng, Nian-Sheng},
  journal={Industrial \& engineering chemistry research},
  volume={47},
  number={9},
  pages={3285--3288},
  year={2008},
  publisher={ACS Publications}
}

@article{takamura_physical_2012,
  title={Physical properties of aqueous glycerol solutions},
  author={Takamura, Koichi and Fischer, Herbert and Morrow, Norman R},
  journal={Journal of Petroleum Science and Engineering},
  volume={98},
  pages={50--60},
  year={2012},
  publisher={Elsevier}
}

@article{dixit_viscoelastic_2025,
  title={Viscoelastic Worthington jets and droplets produced by bursting bubbles},
  author={Dixit, Ayush and Oratis, Alexandros and Zinelis, Konstantinos and Lohse, Detlef and Sanjay, Vatsal},
  journal={Journal of Fluid Mechanics},
  volume={1010},
  pages={A2},
  year={2025},
  publisher={Cambridge University Press}
}

@article{blanco_sea_2020,
  title={On the sea spray aerosol originated from bubble bursting jets},
  author={Blanco--Rodr{\'\i}guez, Francisco J and Gordillo, JM},
  journal={Journal of Fluid Mechanics},
  volume={886},
  pages={R2},
  year={2020},
  publisher={Cambridge University Press}
}

@article{popinet_accurate_2009,
  title={An accurate adaptive solver for surface-tension-driven interfacial flows},
  author={Popinet, St{\'e}phane},
  journal={Journal of Computational Physics},
  volume={228},
  number={16},
  pages={5838--5866},
  year={2009},
  publisher={Elsevier}
}

@article{walls_jet_2015,
  title={Jet drops from bursting bubbles: How gravity and viscosity couple to inhibit droplet production},
  author={Walls, Peter LL and Henaux, Louis and Bird, James C},
  journal={Physical Review E},
  volume={92},
  number={2},
  pages={021002},
  year={2015},
  publisher={APS}
}

@article{brasz_minimum_2018,
  title={Minimum size for the top jet drop from a bursting bubble},
  author={Brasz, C Frederik and Bartlett, Casey T and Walls, Peter LL and Flynn, Elena G and Yu, Yingxian Estella and Bird, James C},
  journal={Physical Review Fluids},
  volume={3},
  number={7},
  pages={074001},
  year={2018},
  publisher={APS}
}

@article{lee_size_2011,
  title={Size limits the formation of liquid jets during bubble bursting},
  author={Lee, Ji San and Weon, Byung Mook and Park, Su Ji and Je, Jung Ho and Fezzaa, Kamel and Lee, Wah-Keat},
  journal={Nature communications},
  volume={2},
  number={1},
  pages={367},
  year={2011},
  publisher={Nature Publishing Group UK London}
}

@article{yang_jet_2025,
  title={Jet Size Prediction in Compound Multiphase Bubble Bursting},
  author={Yang, Zhengyu and Liu, Yang and Feng, Jie},
  journal={Physical Review Letters},
  volume={134},
  number={21},
  pages={214001},
  year={2025},
  publisher={APS}
}

@misc{bubbleShape_h,
  author       = {Berny, A.},
  title        = {bubbleShape.h},
  year         = {2021},
  howpublished = {\url{http://basilisk.fr/sandbox/aberny/bubble/bubbleShape.h}},
}

@article{duchemin_jet_2002,
  title={Jet formation in bubbles bursting at a free surface},
  author={Duchemin, Laurent and Popinet, St{\'e}phane and Josserand, Christophe and Zaleski, St{\'e}phane},
  journal={Physics of fluids},
  volume={14},
  number={9},
  pages={3000--3008},
  year={2002},
  publisher={American Institute of Physics}
}

@article{zeff_singularity_00,
  title={Singularity dynamics in curvature collapse and jet eruption on a fluid surface},
  author={Zeff, Benjamin W and Kleber, Benjamin and Fineberg, Jay and Lathrop, Daniel P},
  journal={Nature},
  volume={403},
  number={6768},
  pages={401--404},
  year={2000},
  publisher={Nature Publishing Group UK London}
}

@article{toba_drop_1959,
  title={Drop production by bursting of air bubbles on the sea surface (II) theoretical study on the shape of floating bubbles},
  author={Toba, Yoshiaki},
  journal={Journal of the Oceanographical Society of Japan},
  volume={15},
  number={3},
  pages={121--130},
  year={1959},
  publisher={The Oceanographic Society of Japan}
}

@article{princen_shape_1963,
  title={Shape of a fluid drop at a liquid-liquid interface},
  author={Princen, HM},
  journal={Journal of Colloid Science},
  volume={18},
  number={2},
  pages={178--195},
  year={1963},
  publisher={Elsevier}
}

@article{sussman_coupled_2000,
  title={A coupled level set and volume-of-fluid method for computing 3D and axisymmetric incompressible two-phase flows},
  author={Sussman, Mark and Puckett, Elbridge Gerry},
  journal={Journal of computational physics},
  volume={162},
  number={2},
  pages={301--337},
  year={2000},
  publisher={Elsevier}
}

@Misc{Basilisk,
	author = {Popinet, S. and collaborators},
	title = {Basilisk},
	howpublished = {\url{http://basilisk.fr}},
	year = {2013--2025}
}

@article{constante_dynamics_2021,
  title={Dynamics of a surfactant-laden bubble bursting through an interface},
  author={Constante-Amores, Cristian R and Kahouadji, Lyes and Batchvarov, Assen and Shin, Seungwon and Chergui, Jalel and Juric, Damir and Matar, Omar K},
  journal={Journal of Fluid Mechanics},
  volume={911},
  pages={A57},
  year={2021},
  publisher={Cambridge University Press}
}

@article{bourouiba_fluid_2021,
  title={The fluid dynamics of disease transmission},
  author={Bourouiba, Lydia},
  journal={Annual Review of Fluid Mechanics},
  volume={53},
  number={1},
  pages={473--508},
  year={2021},
  publisher={Annual Reviews}
}

@article{blanchard_bursting_1954,
  author    = {Blanchard, D. C.},
  title     = {Bursting of Bubbles at an Air--Water Interface},
  journal   = {Nature},
  year      = {1954},
  volume    = {173},
  number    = {4413},
  pages     = {1048},
publisher={Nature Publishing Group UK London}
}

@article{woodcock_giant_1953,
  title={Giant condensation nuclei from bursting bubbles},
  author={Woodcock, A. H. and Kientzler, C. F. and Arons, A. B. and Blanchard, D. C.},
  journal={Nature},
  volume={172},
  number={4390},
  pages={1144--1145},
  year={1953},
  publisher={Nature Publishing Group UK London}
}

@article{blanchard_ejection_1989,
  title={The ejection of drops from the sea and their enrichment with bacteria and other materials: a review},
  author={Blanchard, D. C.},
  journal={Estuaries},
  volume={12},
  pages={127--137},
  year={1989},
  publisher={Springer}
}

@article{rudolph2024bubble,
  title={Bubble ascent and rupture in mud volcanoes},
  author={Rudolph, Maxwell L and Sahu, Kirti Chandra and Savva, Nikos and Szil{\'a}gyi, Andr{\'a}s and H{\'o}rv{\"o}lgyi, Zolt{\'a}n and M{\'a}rton, P{\'e}ter and Tajti, {\'A}d{\'a}m and Sz{\'e}p, K{\'a}roly and Balog, Bogl{\'a}rka and Tripathi, Manoj Kumar and others},
  journal={Royal Society Open Science},
  volume={11},
  number={7},
  pages={231555},
  year={2024},
  publisher={The Royal Society}
}

@article{auregan2025surface,
  title={Surface bubble lifetime in the presence of a turbulent air flow, and the effect of surface layer renewal},
  author={Aur{\'e}gan, Tristan and Deike, Luc},
  journal={arXiv preprint arXiv:2505.04819},
  year={2025}
}

@article{kaganer_structure_1999,
  title = {Structure and Phase Transitions in {{Langmuir}} Monolayers},
  author = {Kaganer, Vladimir M. and M{\"o}hwald, Helmuth and Dutta, Pulak},
  year = {1999},
  month = apr,
  journal = {Reviews of Modern Physics},
  volume = {71},
  number = {3},
  pages = {779--819},
  issn = {0034-6861, 1539-0756},
  doi = {10.1103/RevModPhys.71.779},
  urldate = {2025-09-22},
  copyright = {http://link.aps.org/licenses/aps-default-license},
  langid = {english}
}

@article{berny_role_2020,
  title = {Role of All Jet Drops in Mass Transfer from Bursting Bubbles},
  author = {Berny, Alexis and Deike, Luc and S{\'e}on, Thomas and Popinet, St{\'e}phane},
  year = {2020},
  month = mar,
  journal = {Physical Review Fluids},
  volume = {5},
  number = {3},
  pages = {033605},
  issn = {2469-990X},
  doi = {10.1103/PhysRevFluids.5.033605},
  urldate = {2023-11-29},
  langid = {english}
}

@article{berny_statistics_2021,
  title = {Statistics of {{Jet Drop Production}}},
  author = {Berny, A. and Popinet, S. and S{\'e}on, T. and Deike, L.},
  year = {2021},
  month = may,
  journal = {Geophysical Research Letters},
  volume = {48},
  number = {10},
  pages = {e2021GL092919},
  issn = {0094-8276, 1944-8007},
  doi = {10.1029/2021GL092919},
  urldate = {2023-11-27},
  langid = {english}
}

@article{deike_dynamics_2018,
  title = {Dynamics of Jets Produced by Bursting Bubbles},
  author = {Deike, Luc and Ghabache, Elisabeth and {Liger-Belair}, G{\'e}rard and Das, Arup K. and Zaleski, St{\'e}phane and Popinet, St{\'e}phane and S{\'e}on, Thomas},
  year = {2018},
  month = jan,
  journal = {Physical Review Fluids},
  volume = {3},
  number = {1},
  pages = {013603},
  issn = {2469-990X},
  doi = {10.1103/PhysRevFluids.3.013603},
  urldate = {2025-02-19},
  langid = {english}
}

@article{deike_mass_2022,
  title = {Mass {{Transfer}} at the {{Ocean}}--{{Atmosphere Interface}}: {{The Role}} of {{Wave Breaking}}, {{Droplets}}, and {{Bubbles}}},
  shorttitle = {Mass {{Transfer}} at the {{Ocean}}--{{Atmosphere Interface}}},
  author = {Deike, Luc},
  year = {2022},
  journal = {Annual Review of Fluid Mechanics},
  volume = {54},
  number = {1},
  pages = {191--224},
  doi = {10.1146/annurev-fluid-030121-014132},
  urldate = {2024-01-31}
}

@article{deike_mechanistic_2022,
  title = {A {{Mechanistic Sea Spray Generation Function Based}} on the {{Sea State}} and the {{Physics}} of {{Bubble Bursting}}},
  author = {Deike, L. and Reichl, B. G. and Paulot, F.},
  year = {2022},
  journal = {AGU Advances},
  volume = {3},
  number = {6},
  pages = {e2022AV000750},
  issn = {2576-604X},
  doi = {10.1029/2022AV000750},
  urldate = {2024-03-14},
  copyright = {{\copyright} 2022. The Authors.},
  langid = {english}
}

@article{erinin_comparison_2023,
  title = {Comparison between Shadow Imaging and In-Line Holography for Measuring Droplet Size Distributions},
  author = {Erinin, Martin A. and N{\'e}el, Baptiste and Mazzatenta, Megan T. and Duncan, James H. and Deike, Luc},
  year = {2023},
  month = may,
  journal = {Experiments in Fluids},
  volume = {64},
  number = {5},
  pages = {96},
  issn = {1432-1114},
  doi = {10.1007/s00348-023-03633-8},
  urldate = {2024-11-19},
  langid = {english}
}

@article{erinin_effects_2023,
  title = {The Effects of Surfactants on Plunging Breakers},
  author = {Erinin, M. A. and Liu, C. and Liu, X. and Mostert, W. and Deike, L. and Duncan, J. H.},
  year = {2023},
  month = oct,
  journal = {Journal of Fluid Mechanics},
  volume = {972},
  pages = {R5},
  issn = {0022-1120, 1469-7645},
  doi = {10.1017/jfm.2023.721},
  urldate = {2024-03-18},
  langid = {english}
}

@article{fainerman_adsorption_2009,
  title = {Adsorption Layer Characteristics of {{Triton}} Surfactants {{Part}} 2. {{Dynamic}} surface tension and adsorption},
  author = {Fainerman, V.B. and Lylyk, S.V. and Aksenenko, E.V. and Liggieri, L. and Makievski, A.V. and Petkov, J.T. and Yorke, J. and Miller, R.},
  year = {2009},
  month = feb,
  journal = {Colloids and Surfaces A: Physicochemical and Engineering Aspects},
  volume = {334},
  number = {1-3},
  pages = {8--15},
  issn = {09277757},
  doi = {10.1016/j.colsurfa.2008.09.052},
  urldate = {2025-04-18},
  copyright = {https://www.elsevier.com/tdm/userlicense/1.0/},
  langid = {english}
}

@article{fainerman_adsorption_2009a,
  title = {Adsorption Layer Characteristics of {{Triton}} Surfactants 1. {{Surface}} tension and adsorption isotherms},
  author = {Fainerman, V.B. and Lylyk, S.V. and Aksenenko, E.V. and Makievski, A.V. and Petkov, J.T. and Yorke, J. and Miller, R.},
  year = {2009},
  month = feb,
  journal = {Colloids and Surfaces A: Physicochemical and Engineering Aspects},
  volume = {334},
  number = {1-3},
  pages = {1--7},
  issn = {09277757},
  doi = {10.1016/j.colsurfa.2008.09.015},
  urldate = {2025-04-18},
  copyright = {https://www.elsevier.com/tdm/userlicense/1.0/},
  langid = {english}
}

@article{farsoiya_coupled_2024,
  title = {Coupled Volume of Fluid and Phase Field Method for Direct Numerical Simulation of Insoluble Surfactant-Laden Interfacial Flows and Application to Rising Bubbles},
  author = {Farsoiya, Palas Kumar and Popinet, St{\'e}phane and Stone, Howard A. and Deike, Luc},
  year = {2024},
  month = sep,
  journal = {Physical Review Fluids},
  volume = {9},
  number = {9},
  pages = {094004},
  publisher = {American Physical Society},
  doi = {10.1103/PhysRevFluids.9.094004},
  urldate = {2024-12-30}
}

@article{ganan-calvo_revision_2017,
  title = {Revision of {{Bubble Bursting}}: {{Universal Scaling Laws}} of {{Top Jet Drop Size}} and {{Speed}}},
  shorttitle = {Revision of {{Bubble Bursting}}},
  author = {{Ga{\~n}{\'a}n-Calvo}, Alfonso M.},
  year = {2017},
  month = nov,
  journal = {Physical Review Letters},
  volume = {119},
  number = {20},
  pages = {204502},
  publisher = {American Physical Society},
  doi = {10.1103/PhysRevLett.119.204502},
  urldate = {2025-03-20}
}

@article{ganan-calvo_scaling_2018,
  title = {Scaling Laws of Top Jet Drop Size and Speed from Bubble Bursting Including Gravity and Inviscid Limit},
  author = {{Ga{\~n}{\'a}n-Calvo}, Alfonso M.},
  year = {2018},
  month = sep,
  journal = {Physical Review Fluids},
  volume = {3},
  number = {9},
  pages = {091601},
  issn = {2469-990X},
  doi = {10.1103/PhysRevFluids.3.091601},
  urldate = {2025-04-03},
  langid = {english}
}

@article{ghabache_physics_2014,
  title = {On the Physics of Fizziness: {{How}} Bubble Bursting Controls Droplets Ejection},
  shorttitle = {On the Physics of Fizziness},
  author = {Ghabache, Elisabeth and Antkowiak, Arnaud and Josserand, Christophe and S{\'e}on, Thomas},
  year = {2014},
  month = dec,
  journal = {Physics of Fluids},
  volume = {26},
  number = {12},
  pages = {121701},
  issn = {1070-6631},
  doi = {10.1063/1.4902820},
  urldate = {2025-05-02}
}

@article{gordillo_capillary_2019,
  title = {Capillary Waves Control the Ejection of Bubble Bursting Jets},
  author = {Gordillo, J. M. and {Rodr{\'i}guez-Rodr{\'i}guez}, J.},
  year = {2019},
  month = may,
  journal = {Journal of Fluid Mechanics},
  volume = {867},
  pages = {556--571},
  issn = {0022-1120, 1469-7645},
  doi = {10.1017/jfm.2019.161},
  urldate = {2025-03-24},
  langid = {english}
}

@article{ji_secondary_2023a,
  title = {Secondary {{Bubble Entrainment}} via {{Primary Bubble Bursting}} at a {{Viscoelastic Surface}}},
  author = {Ji, Bingqiang and Yang, Zhengyu and Wang, Zirui and Ewoldt, Randy H. and Feng, Jie},
  year = {2023},
  month = sep,
  journal = {Physical Review Letters},
  volume = {131},
  number = {10},
  pages = {104002},
  issn = {0031-9007, 1079-7114},
  doi = {10.1103/PhysRevLett.131.104002},
  urldate = {2025-04-29},
  langid = {english}
}

@article{jiang_submicron_2022,
  title = {Submicron Drops from Flapping Bursting Bubbles},
  author = {Jiang, Xinghua and Rotily, Lucas and Villermaux, Emmanuel and Wang, Xiaofei},
  year = {2022},
  month = jan,
  journal = {Proceedings of the National Academy of Sciences},
  volume = {119},
  number = {1},
  pages = {e2112924119},
  issn = {0027-8424, 1091-6490},
  doi = {10.1073/pnas.2112924119},
  urldate = {2023-12-04},
  langid = {english}
}

@article{lai_bubble_2018,
  title = {Bubble {{Bursting}}: {{Universal Cavity}} and {{Jet Profiles}}},
  shorttitle = {Bubble {{Bursting}}},
  author = {Lai, Ching-Yao and Eggers, Jens and Deike, Luc},
  year = {2018},
  month = oct,
  journal = {Physical Review Letters},
  volume = {121},
  number = {14},
  pages = {144501},
  issn = {0031-9007, 1079-7114},
  doi = {10.1103/PhysRevLett.121.144501},
  urldate = {2025-04-02},
  langid = {english}
}

@article{liao_deformation_2006,
  title = {Deformation and Breakup of a Stretching Liquid Bridge Covered with an Insoluble Surfactant Monolayer},
  author = {Liao, Ying-Chih and Franses, Elias I. and Basaran, Osman A.},
  year = {2006},
  month = feb,
  journal = {Physics of Fluids},
  volume = {18},
  number = {2},
  pages = {022101},
  issn = {1070-6631, 1089-7666},
  doi = {10.1063/1.2166657},
  urldate = {2024-08-27},
  langid = {english}
}

@article{manikantan_surfactant_2020,
  title = {Surfactant Dynamics: Hidden Variables Controlling Fluid Flows},
  shorttitle = {Surfactant Dynamics},
  author = {Manikantan, Harishankar and Squires, Todd M.},
  year = {2020},
  month = jun,
  journal = {Journal of Fluid Mechanics},
  volume = {892},
  pages = {P1},
  issn = {0022-1120, 1469-7645},
  doi = {10.1017/jfm.2020.170},
  urldate = {2024-05-24},
  langid = {english}
}

@article{mazzatenta_linking_2025,
  title={Linking emitted drops to collective bursting bubbles across a wide range of bubble size distributions},
  author={Mazzatenta, Megan and Erinin, Martin A and N{\'e}el, Baptiste and Deike, Luc},
  journal={Journal of Fluid Mechanics},
  volume={1015},
  pages={A8},
  year={2025},
  publisher={Cambridge University Press}
}

@article{miller_dynamic_2017,
  title = {Dynamic Interfacial Tension of Surfactant Solutions},
  author = {Miller, R. and Aksenenko, E.V. and Fainerman, V.B.},
  year = {2017},
  month = sep,
  journal = {Advances in Colloid and Interface Science},
  volume = {247},
  pages = {115--129},
  issn = {00018686},
  doi = {10.1016/j.cis.2016.12.007},
  urldate = {2025-01-30},
  langid = {english}
}

@article{neel_role_2022,
  title = {Role of {{Contamination}} in {{Optimal Droplet Production}} by {{Collective Bubble Bursting}}},
  author = {N{\'e}el, B. and Erinin, M. A. and Deike, L.},
  year = {2022},
  month = jan,
  journal = {Geophysical Research Letters},
  volume = {49},
  number = {1},
  pages = {e2021GL096740},
  issn = {0094-8276, 1944-8007},
  doi = {10.1029/2021GL096740},
  urldate = {2023-11-27},
  langid = {english}
}

@article{pico_surfactantladen_2024,
  title = {Surfactant-Laden Bubble Bursting: {{Dynamics}} of Capillary Waves and {{Worthington}} Jet at Large {{Bond}} Number},
  shorttitle = {Surfactant-Laden Bubble Bursting},
  author = {Pico, P. and Kahouadji, L. and Shin, S. and Chergui, J. and Juric, D. and Matar, O. K.},
  year = {2024},
  month = aug,
  journal = {Physical Review Fluids},
  volume = {9},
  number = {8},
  pages = {083606},
  issn = {2469-990X},
  doi = {10.1103/PhysRevFluids.9.083606},
  urldate = {2025-02-17},
  langid = {english}
}

@article{pierre_influence_2022,
  title = {Influence of Surfactant Concentration on Drop Production by Bubble Bursting},
  author = {Pierre, Juliette and Poujol, Mathis and S{\'e}on, Thomas},
  year = {2022},
  month = jul,
  journal = {Physical Review Fluids},
  volume = {7},
  number = {7},
  pages = {073602},
  publisher = {American Physical Society},
  doi = {10.1103/PhysRevFluids.7.073602},
  urldate = {2024-03-14}
}

@article{poulain_ageing_2018,
  title = {Ageing and Burst of Surface Bubbles},
  author = {Poulain, S. and Villermaux, E. and Bourouiba, L.},
  year = {2018},
  month = sep,
  journal = {Journal of Fluid Mechanics},
  volume = {851},
  pages = {636--671},
  issn = {0022-1120, 1469-7645},
  doi = {10.1017/jfm.2018.471},
  urldate = {2023-12-12},
  langid = {english}
}

@article{poulain_biosurfactants_2018,
  title = {Biosurfactants {{Change}} the {{Thinning}} of {{Contaminated Bubbles}} at {{Bacteria-Laden Water Interfaces}}},
  author = {Poulain, S. and Bourouiba, L.},
  year = {2018},
  month = nov,
  journal = {Physical Review Letters},
  volume = {121},
  number = {20},
  pages = {204502},
  issn = {0031-9007, 1079-7114},
  doi = {10.1103/PhysRevLett.121.204502},
  urldate = {2024-03-14},
  langid = {english}
}

@article{shaw_film_2024,
  title = {Film Drop Production over a Wide Range of Liquid Conditions},
  author = {Shaw, Daniel B. and Deike, Luc},
  year = {2024},
  month = mar,
  journal = {Physical Review Fluids},
  volume = {9},
  number = {3},
  pages = {033602},
  issn = {2469-990X},
  doi = {10.1103/PhysRevFluids.9.033602},
  urldate = {2024-03-29},
  langid = {english}
}

@article{sierou_selfsimilar_2004,
  title = {Self-Similar Recoil of Inviscid Drops},
  author = {Sierou, Asimina and Lister, John R.},
  year = {2004},
  month = may,
  journal = {Physics of Fluids},
  volume = {16},
  number = {5},
  pages = {1379--1394},
  issn = {1070-6631, 1089-7666},
  doi = {10.1063/1.1689031},
  urldate = {2025-04-25},
  langid = {english}
}

@article{vega_influence_2024,
  title = {Influence of a Surfactant on Bubble Bursting},
  author = {Vega, E.J. and Montanero, J.M.},
  year = {2024},
  month = feb,
  journal = {Experimental Thermal and Fluid Science},
  volume = {151},
  pages = {111097},
  issn = {08941777},
  doi = {10.1016/j.expthermflusci.2023.111097},
  urldate = {2025-02-17},
  langid = {english}
}

@article{veron_ocean_2015,
  title = {Ocean {{Spray}}},
  author = {Veron, Fabrice},
  year = {2015},
  month = jan,
  journal = {Annual Review of Fluid Mechanics},
  volume = {47},
  number = {Volume 47, 2015},
  pages = {507--538},
  publisher = {Annual Reviews},
  issn = {0066-4189, 1545-4479},
  doi = {10.1146/annurev-fluid-010814-014651},
  urldate = {2024-06-03},
  langid = {english}
}

@article{wang_role_2017,
  title = {The Role of Jet and Film Drops in Controlling the Mixing State of Submicron Sea Spray Aerosol Particles},
  author = {Wang, Xiaofei and Deane, Grant B. and Moore, Kathryn A. and Ryder, Olivia S. and Stokes, M. Dale and Beall, Charlotte M. and Collins, Douglas B. and Santander, Mitchell V. and Burrows, Susannah M. and Sultana, Camille M. and Prather, Kimberly A.},
  year = {2017},
  month = jul,
  journal = {Proceedings of the National Academy of Sciences},
  volume = {114},
  number = {27},
  pages = {6978--6983},
  publisher = {Proceedings of the National Academy of Sciences},
  doi = {10.1073/pnas.1702420114},
  urldate = {2025-04-14}
}

@article{yang_enhanced_2023a,
  title = {Enhanced Singular Jet Formation in Oil-Coated Bubble Bursting},
  author = {Yang, Zhengyu and Ji, Bingqiang and Ault, Jesse T. and Feng, Jie},
  year = {2023},
  month = jun,
  journal = {Nature Physics},
  volume = {19},
  number = {6},
  pages = {884--890},
  issn = {1745-2473, 1745-2481},
  doi = {10.1038/s41567-023-01958-z},
  urldate = {2025-04-17},
  langid = {english}
}

@article{gaver_dynamics_1990,
  title={The dynamics of a localized surfactant on a thin film},
  author={Gaver, Donald P and Grotberg, James B},
  journal={Journal of Fluid Mechanics},
  volume={213},
  pages={127--148},
  year={1990},
  publisher={Cambridge University Press}
}

@article{quere_fluid_1999,
  title={Fluid coating on a fiber},
  author={Qu{\'e}r{\'e}, David},
  journal={Annual review of fluid mechanics},
  volume={31},
  number={1},
  pages={347--384},
  year={1999},
  publisher={Annual Reviews 4139 El Camino Way, PO Box 10139, Palo Alto, CA 94303-0139, USA}
}

@article{rodriguez_critical_2025,
  title={Critical bubble bursting in real water. Effect of surface-active contaminants},
  author={Rodr{\'\i}guez-Aparicio, S and Cebri{\'a}n-Garc{\'\i}a, A. and Vega, E. J. and Montanero, J. M. and Cabezas, M. G.},
  journal={arXiv preprint arXiv:2507.17671},
  year={2025}
}

@article{ganan_ocean_2023,
  title={The ocean fine spray},
  author={Ga{\~n}{\'a}n-Calvo, Alfonso M},
  journal={Physics of Fluids},
  volume={35},
  number={2},
  year={2023},
  publisher={AIP Publishing}
}

@article{saini_implementation_2025,
title = {Implementation of integral surface tension formulations in a volume of fluid framework and their applications to Marangoni flows},
journal = {Journal of Computational Physics},
volume = {542},
pages = {114348},
year = {2025},
issn = {0021-9991},
doi = {https://doi.org/10.1016/j.jcp.2025.114348},
url = {https://www.sciencedirect.com/science/article/pii/S0021999125006308},
author = {Mandeep Saini and Vatsal Sanjay and Youssef Saade and Detlef Lohse and Stéphane Popinet},
}

@misc{eshima_code_2025,
  author       = {Eshima, Jun and Aur{\'e}gan, Tristan and Farsoiya, Palas Kumar and Popinet, St{\'e}phane and Stone, Howard A and Deike, Luc},
  title        = {Code for ``Size Amplification of Jet drops due to Insoluble Surfactants"},
  howpublished = {\url{https://basilisk.fr/sandbox/jeshima/jet_drop_surfactant_effect/}},
  year         = {2025}
}

@electronic{eshima_data_2025,
  author      = {Eshima, Jun and Aur{\'e}gan, Tristan and Farsoiya, Palas Kumar and Popinet, St{\'e}phane and Stone, Howard A and Deike, Luc},
  title       = {{Data set for "Size Amplification of Jet Drops due to Insoluble Surfactants"}},
  version     = 1,
  publisher   = {{Princeton University}},
  year        = 2025,
  url         = {https://doi.org/10.34770/1pf8-xy34}
}

\end{document}